\documentclass[article,superscriptaddress]{revtex4-1}
\usepackage{multirow}
\usepackage[table]{xcolor}
\usepackage{rotating}
\usepackage{amsmath}
\usepackage{bm}
\usepackage{array}
\usepackage{graphicx}
\usepackage{dcolumn}   
\usepackage{amssymb}
\usepackage[english]{babel}
\usepackage[colorlinks=true, citecolor=red, linkcolor=blue,urlcolor=blue ]{hyperref}
\def\v{\vspace{2cm}}

\makeatletter
\renewcommand{\p@subsection}{}
\renewcommand{\p@subsubsection}{}
\makeatother

\begin{document}
\title{The swimming of a deforming helix}
\date{\today}
\author{Lyndon Koens\footnote{Present address: Department of Mathematics and Statistics, Macquarie University, 2113, NSW, Australia.
\\ Email: lyndon.koens@mq.edu.au}}
\affiliation{ Department of Applied Mathematics and Theoretical Physics, University of Cambridge, Wilberforce Road, Cambridge CB3 0WA, United Kingdom}
\author{Hang Zhang} \thanks{Present address:  Department of Applied Physics,  Aalto University, Puumiehenkuja 2 02150 Espoo, Finland}
\affiliation{DWI-Leibniz Institute for Interactive Materials
RWTH Aachen University, Forckenbeck str.~50, D-52056 Aachen, Germany}
\author{Martin Moeller}
\affiliation{DWI-Leibniz Institute for Interactive Materials
RWTH Aachen University, Forckenbeck str.~50, D-52056 Aachen, Germany}
\author{Ahmed Mourran}
\affiliation{DWI-Leibniz Institute for Interactive Materials
RWTH Aachen University, Forckenbeck str.~50, D-52056 Aachen, Germany}
\author{Eric Lauga}
\email{e.lauga@damtp.cam.ac.uk}
\affiliation{ Department of Applied Mathematics and Theoretical Physics, University of Cambridge, Wilberforce Road, Cambridge CB3 0WA, United Kingdom}
\begin{abstract} 
Many microorganisms and artificial microswimmers use helical appendages in order to generate locomotion. Though often rotated so as to produce thrust, some species of bacteria such \textit{Spiroplasma},  \textit{Rhodobacter sphaeroides} and \textit{Spirochetes} induce movement by deforming a helical-shaped 
body. Recently, artificial devices have been created which also generate motion by deforming their helical body in a non-reciprocal way (Mourran \textit{et al.}~Adv. Mater.~{\bf 29}, 1604825, 2017). Inspired by these systems, we investigate the transport of a deforming helix within a viscous fluid.  Specifically, we consider a  swimmer that maintains a helical centreline and a single handedness while changing its helix radius, pitch and wavelength uniformly across the body. We first discuss how a deforming helix can create a non-reciprocal translational and rotational swimming stroke and identify its principle direction of motion. We then determine the leading-order physics for helices with small helix radius before considering the general behaviour for different configuration parameters and how these swimmers can be optimised. Finally, we explore how the presence of walls, gravity, and defects in the centreline allow the helical device to break symmetries, increase its speed, and generate transport in directions not available to helices in bulk fluids.
\end{abstract}
\maketitle
\def\v{\vspace{2cm}}

\section{Introduction}

In the early 1950s, GI Taylor offered the first fluid mechanical explanation of how microscopic organisms are able to swim at low Reynolds number \cite{Taylor1951}. Since then, decades of close collaborations between experimentalists and theorists have greatly improved our knowledge of fluid-based motion in the microscopic world \cite{Lauga2009,Elgeti2015,Gaffney2011,Lauga2016}. We now understand how, at these small scales, the creation of net movement critically depends on the anisotropy of the viscous drag within the fluid \cite{BECKER2003}, and requires a non-reciprocal swimming stroke to break the time-reversal symmetry of the underlying equations of motion \cite{Purcell}. The motion of model cellular swimmers, such as spermatozoa, bacteria, and  algae,  have been studied at length in order to characterise how they swim in bulk fluids \cite{Turner2000, Chattopadhyay2006, Gaffney2011, Goldstein2015, Hu2015, Adhyapak2016, Hintsche2017, Riley2018}, find food sources \cite{Locsei2007,Locsei2009,Drescher2010, Buchmann2018}, behave near boundaries \cite{Spagnolie2012,Denissenko2012a,Shum2015a,Bianchi2017, Kuhn2017} and move in complex fluids \cite{Ullrich2016, Balin2017,Zottl2017,Ho2018}. 

These investigations have also prompted the creation of artificial microswimmers. While some synthetic devices have been designed to prove theoretical models \cite{Lumay2013,Grosjean2015} or to exploit propulsion mechanisms for rigid shapes \cite{Walther2008, Mangal2017}, many artificial swimmers are directly inspired by  propulsion methods used in the biological world \cite{Zhang2010, Tottori2012, Tottori2013, Diller2014, Xu2015a, Mourran2017, Sayyaadi2017, Ali2017}. Two popular biological methodologies to induce  motion at small scales are the planar waving of slender filaments, commonly used by spermatozoa \cite{Gaffney2011, Diller2014}, or the rotating of semi-rigid helical structures, commonly used by bacteria   \cite{Chattopadhyay2006, Zhang2010, Lauga2016}.

Though the rotation of rigid helices is associated classically with swimming at low Reynolds numbers, the deformation of a helix can also produce motion \cite{Shaevitz2005, Xu2015a}. The helical flagellar filaments of \textit{Escherichia coli} (\textit{E. coli}), for example, are polymorphic \cite{Calldine1978, Calladine2013} and can switch forms when rotated in a different direction \cite{Wada2008, Vogel2010, Adhyapak2016, Ko2017}, leading to random reorientations of the bacteria as a whole \cite{Berg1993book, Lauga2016}. For the bacterium \textit{Rhodobacter sphaeroides}, these reorientation events were previously thought to be governed by Brownian rotation, but recent results suggest that the polymorphic change itself actively rotates the swimmer \cite{Rosser2014a}. 

A different phylum of bacteria, the \textit{Spirochaetes}, can also use the deformation of helix to generate propulsion. These bacteria encase their flagellar filaments within a thin sheath around their body \cite{Goldstein1990, Li2000, Koens2014}. When the helical filaments are rotated, their motion deforms the cell body through a set of waves and helical shapes, that in turn creates motion \cite{ Kan2007, Jung2007, Dombrowski2009}. Similarly \textit{Spiroplasma}, a group of small helical unflagellated bacteria, are able to generate thrust by propagating a switch in helix handedness along its body length \cite{Shaevitz2005,Wada2007, Yang2009}.

\begin{figure} [t]
\centering
\includegraphics[width=0.8\textwidth]{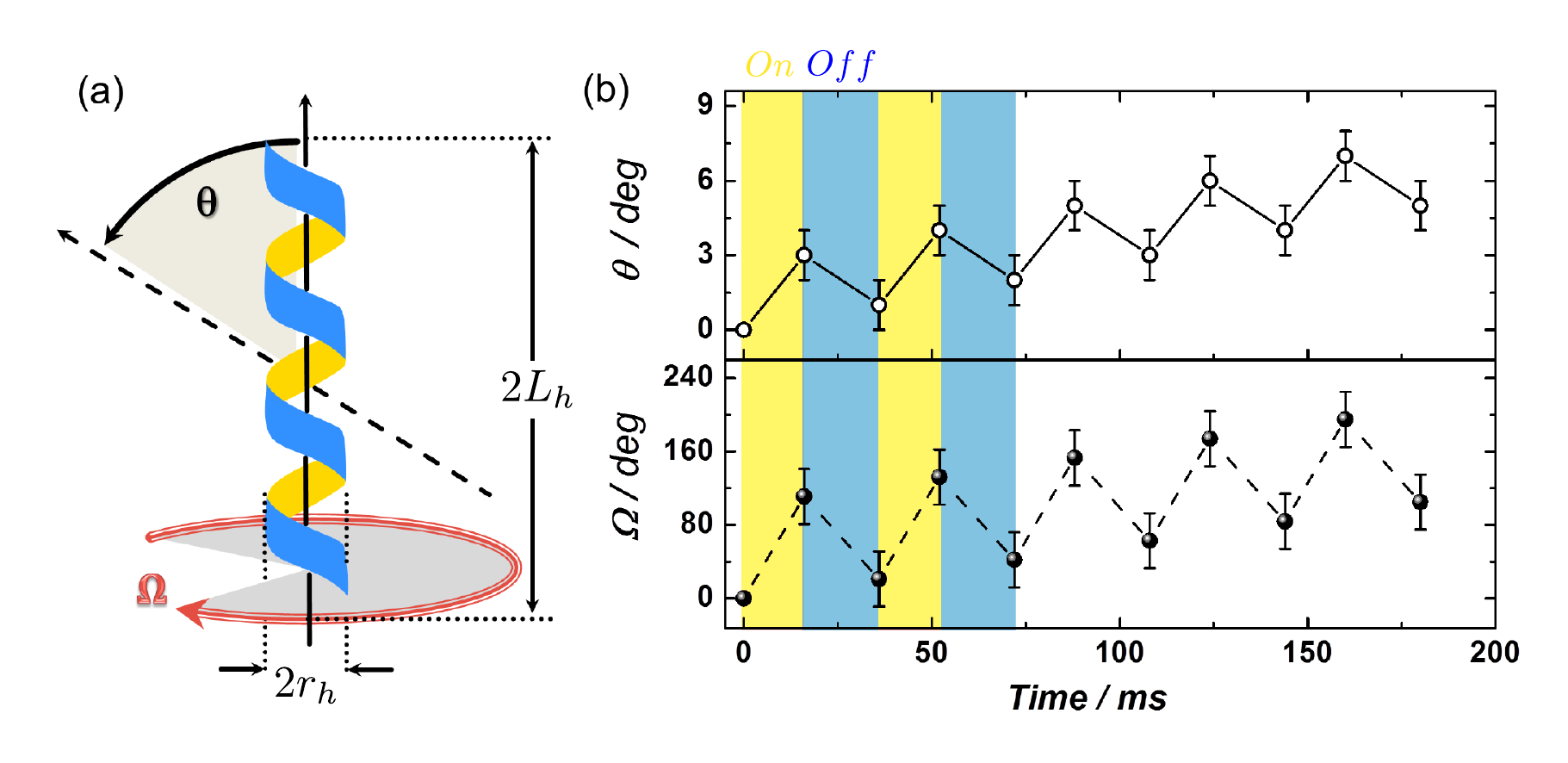}
\caption{Net rotating motion of a  deforming gel helix \cite{Mourran2017}; 
(a) Schematic diagram of the helix of length 
$2 L_{h}$ (measured along the helix axis) and diameter $2 r_{h}$, with a definition of the two angles $(\theta, \Omega)$ by which it may rotate; (b) Angular displacements of the helix as it periodically deforms in a non-reciprocal manner. Light yellow regions indicate when the light is on, while dark blue regions reflect when the light is off. Figure adapted from Ref.~\cite{Mourran2017}  under the Creative Commons License.}
\label{fig:gelswimmer}
\end{figure}

Recently Mourran \textit{et al.}~developed a novel deforming helical microswimmer composed of a temperature sensitive gel~\cite{Mourran2017, Zhang2017}. When periodically heated and cooled with a laser, these gels retain a helical shape while changing their helix radius and axial length in a non-reciprocal manner. These non-reciprocal deformations lead to a net rotation parallel and perpendicular to its helix axis and, when near walls, a net translation (Fig.~\ref{fig:gelswimmer}). 
 
Motivated by these recent experimental results and by the common occurrence of deforming helices in biological systems, this paper explores how a deforming helix can generate motion. First we consider the dynamics of    an inextensible helix  that changes its axial length, helix radius and wavelength in time in order to move in an infinite fluid. We demonstrate that this simple shape can generate a non-reciprocal swimming stroke and that, in an infinite fluid, it may translate and rotate in only one direction by symmetry. The physics of this motion is elaborated upon in the limit of small helix radius before we determine the general trends of these swimmers and the optimal configuration loops. Finally, we consider how the presence of gravity, walls or imperfections in the helical shape can be used to break symmetries and to allow the deforming helix to move in any direction.
 
This article is organised as follows. In sect.~\ref{sec:Kinematics} we describe the configuration of the helix in an infinite fluid, its deformation velocity, symmetries, and the low Reynolds number hydrodynamic model used to calculate its motion. In sect.~\ref{sec:smallamp} we then use this model to determine the dynamics of the deforming helix in the limit of small helix radius and demonstrate that the leading-order motion is very sensitive to the path taken in configuration space. The general behaviour of the swimmer and its optimal configuration loops are then explored in sect.~\ref{sec:full}. Finally in sect.~\ref{sec:symmetry} we consider how gravity, walls and imperfections in the helix can break the symmetry of the swimmer and determine the leading order dynamics when the helix radius is small.

 
\section{Kinematics and dynamics} \label{sec:Kinematics}

Although helices have long been associated with swimming at low Reynolds number, the physics governing the swimming of a deforming  helix has not yet been addressed. This can, however, be achieved through simple considerations of the helical shape and its deformation. In this section we set up the problem by mathematically describing the shape of a deforming helix, the symmetries of its motion, and the hydrodynamic modelling which will be exploited to quantify the motion.

 \subsection{Helical geometry and kinematics}

\begin{figure}[t]
\includegraphics[width=\textwidth]{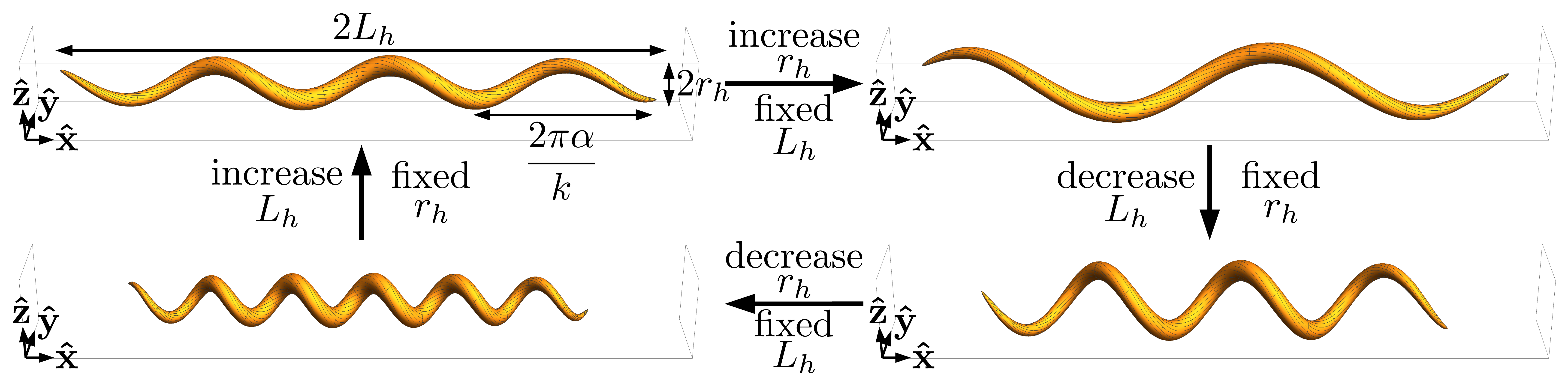}
\caption{Illustration of a helix undergoing a non-reciprocal deformation with a centreline described by Eq.~\eqref{centrerline1} and a circular cross-section. In this example, either the helix radius or the length along its helix axis is changed while the other is fixed. The wavenumber is set through the inextensibility condition $\alpha^{2} +k^{2} r_{h}^{2}=1$ where $\alpha = L_{h}/ \ell$ and $2\ell$ is the total length of the helix centreline.}
\label{fig:cent}
\end{figure}

The deforming helix varies its helix angle, helix radius and wavenumber uniformly across its length to generate motion. As such its centreline, $\mathbf{r}(s,t)$, is a helix at all times and so can be described parametrically as
\begin{equation}
\mathbf{r}(s,t) = \{ \alpha(t) s, r_{h}(t) \cos(k(t) s), r_{h}(t) \sin(k(t) s) \}, \label{centrerline1}
\end{equation}
in the Cartesian coordinate system $\{x,y,z\}$ (Fig.~\ref{fig:cent}). In the above equation, $t$ is time,  $\alpha(t)$ is the cosine of the helix angle, $r_{h}(t)$ is the helix radius, $k(t)$ is the wavenumber, $s \in [\ell,-\ell]$ is the arc length and $2 \ell$ is the total length of the helix along its centreline. The surface of a deforming helix, with a circular cross-section, is then given by
\begin{equation}
\mathbf{S}(s,\Theta,t) = \mathbf{r}(s,t) + r_{f}(s) \mathbf{\hat{e}}_{\rho}(s,\Theta,t), \label{surface}
\end{equation}
where $r_{f}(s)$ is the radius of the filament, $\mathbf{\hat{e}}_{\rho}(s,\Theta,t)$ is the radial unit vector perpendicular to the centreline tangent $\mathbf{\hat{t}}(s,t)= {\partial_s \mathbf{r}}(s,t)$, and $\Theta \in [-\pi,\pi]$ is the azimuthal coordinate of the surface. In this configuration, the length along the helix axis is given by $2 L_{h} (t) = 2 \alpha(t) \ell$.  For an inextensible helix,  $s$ must describe the conserved arclength of the curve for all $t$. This is equivalent to enforcing the derivative of the curve with respect to s to be the centreline tangent, $\mathbf{\hat{t}}(s,t)=\partial_{s} \mathbf{r}(s,t)$. The inextensible constraint can therefore be written as $[\partial_{s} \mathbf{r}(s,t)]^{2} =1$ or
\begin{equation}
\alpha(t)^{2} + r_{h}(t)^{2} k(t)^{2} =1, \label{inext}
\end{equation}
at all times. In what follows, we scale all lengths in the problem by $\ell$.

The centreline, Eq.~\eqref{centrerline1}, and inextensibility constraint, Eq.~\eqref{inext}, show that a helix is uniquely defined by any two of the three conformation parameters: $\alpha$, $r_{h}$, and $k$. A deforming helix therefore has two independent degrees of freedom which can be varied to generate motion. As famously discussed by Purcell in his seminal article on locomotion at low Reynolds numbers \cite{Purcell}, two independent degrees of freedom are sufficient to create a non-reciprocal stroke (\textit{i.e.}~not identical under a time-reversal symmetry) and thus are the minimal requirements needed to induce net motion. An example non-reciprocal stroke, generated by a deforming helix, is shown in Fig.~\ref{fig:cent}. This stroke was obtained by alternatively varying the axial length, $L_{h}= \alpha(t) \ell$, and helix radius, $r_{h}$, while keeping the other parameter fixed.

The surface velocity, due to the deformation of the helix, is given by $\mathbf{V}=\partial_{t} \mathbf{S}(s,\Theta,t)$. In the limit that $r_{f} \ll \ell$ this velocity simplifies to $\partial_{t} \mathbf{S}(s,\Theta,t) \approx \partial_{t} \mathbf{r}(s,t)$. This so-called slender-body limit   is often sufficient to capture the leading physics of many microscopic swimming systems \cite{GRAY1955, 1976, Lauga2009, Koens2014}.  Hence, assuming that our helices are slender, we can approximate the deformation surface velocity as
\begin{eqnarray}
\mathbf{V}(s,t) &\approx &\left\{ \frac{d\alpha}{dt} s,  \frac{d r_{h}}{dt} \cos(k(t) s) - r_{h}(t) s \frac{d k}{dt} \sin(k(t) s),\frac{d r_{h}}{dt} \sin(k(t) s) +r_{h}(t) s \frac{d k}{dt} \cos(k(t) s)\right\} \notag \\
&=&s \left(r_{h}^{2}(t) k(t) \frac{d k}{dt} + \alpha(t) \frac{d\alpha}{dt} \right) \mathbf{\hat{t}} - \frac{d r_{h}}{dt}\mathbf{\hat{n}} - s r_{h} \left( \alpha(t) \frac{d k}{dt} -k(t) \frac{d\alpha}{dt}\right) \mathbf{\hat{b}}, \label{defvel1}
\end{eqnarray}
where the tangent, $\mathbf{\hat{t}}$, normal, $\mathbf{\hat{n}}$, and bi-normal, $\mathbf{\hat{b}}$, vectors to the centreline are 
\begin{eqnarray}
\mathbf{\hat{t}} &=&\{ \alpha(t) , - r_{h}(t) k(t) \sin(k(t) s), r_{h}(t) k(t) \cos(k(t) s) \}, \\
\mathbf{\hat{n}} &=&\{ 0 , - \cos(k(t) s), - \sin(k(t) s) \}, \\
\mathbf{\hat{b}} &=&\{  r_{h}(t) k(t) ,  \alpha(t) \sin(k(t) s) , -\alpha(t) \cos(k(t) s)\}.
\end{eqnarray}
When the helix is located in a fluid, the deformation kinematics in Eq.~\eqref{defvel1} creates  hydrodynamic forces and torques on the body which must balance the viscous drag from the swimming motion of the helix.

\subsection{Symmetry conditions}

Due to the linearity of the low Reynolds number hydrodynamic equations (the Stokes equations), the swimming motion of the deforming helix is subject to the same symmetries as the helical shape, Eq.~\eqref{centrerline1}, and its deformation velocity, Eq.~\eqref{defvel1}. The deforming helix, when located in an infinite fluid, has one such symmetry; namely a $\pi$ rotation symmetry around a director perpendicular to the helix axis. In our parametrisation of the helix this symmetry axis coincides with $\mathbf{\hat{y}}=\{0,1,0\}$.   If, at any time, a helix is rotated by $\pi$ around the symmetry axis the system is identical to before it was rotated. Therefore any net motion perpendicular to $\mathbf{\hat{y}}$, in our parametrisation, must be equal to its negative after this rotation and is thus zero. Hence the deforming helical swimmers only generate force, or motion along $\mathbf{\hat{y}}$. 
 Note as the experimental gel helical swimmer rotates around both the helix axis and perpendicular to it \cite{Mourran2017, Zhang2017}, this rotation symmetry must be broken. In principle this  could be caused by deviations in the shape of the helix, or the presence of walls. Such influences will be discussed further in sect.~\ref{sec:symmetry}.

\subsection{Hydrodynamics}

 The net translation and rotation of the deforming helical swimmer is governed by its interaction with the surrounding viscous fluid. The details of these hydrodynamic interactions dictate how the forces and torques on the body relate to its shape and velocity. For long slender swimmers in viscous environments, many semi-analytical approaches exist to compute the distribution of hydrodynamic forces  \cite{GRAY1955, 1976, Keller1976a, Johnson1979, Koens2016, Montenegro-Johnson2016}. These techniques fall broadly into two classes: slender body theories (SBT) \cite{1976, Keller1976a, Johnson1979} and resistive force theories (RFT) \cite{GRAY1955}. While SBTs relate the local force on the body to the surface velocity through integral equations (which in general have to be inverted numerically), RFTs relate the local force to the local velocity at that point through an anisotropic drag matrix (and as a result can be evaluated analytically).  
The former method is typically accurate to order $r_{f}/\ell$ while the latter is accurate to order $1/\log(r_{f}/\ell)$. The linear relationship between the local force and velocity in RFT is known to capture much of the governing physics  and the qualitative behaviour of the filament. Hence, as RFT is analytical, it is a very useful method to explore and optimise the dynamics of a filamentous swimmer. We will therefore use RFT to describe the hydrodynamic forces of the deforming helix swimmer. Specifically, for a slender body in a quiescent unbounded fluid, the relationship between the local velocity of the body centreline, $\mathbf{U}(s)$, and the hydrodynamic force per unit length acting from the body on the fluid, $\mathbf{f}(s)$, is given by
 \begin{equation} \label{gelRFT}
\mathbf{f}(s) = \left[\zeta_{\parallel} \mathbf{\hat{t}} \mathbf{\hat{t}} + \zeta_{\perp} (\mathbf{I} - \mathbf{\hat{t}} \mathbf{\hat{t}}) \right] \cdot \mathbf{U}(s),
\end{equation}
where $\zeta_{\parallel}$ and $\zeta_{\perp}$ are the drag coefficients for motion parallel and perpendicular to the filaments tangent, respectively. The  drag relationship in Eq.~\eqref{gelRFT} is anisotropic when $\zeta_{\parallel}\neq\zeta_{\perp}$, and for very slender bodies in an unbounded low Reynolds numbers flow  becomes  approximately
 $\zeta_{\perp} \approx 2 \zeta_{\parallel}$  \cite{GRAY1955, Taylor1967, Lauga2009}.
 
 The total hydrodynamic force, $\mathbf{F}(t)$, and torque, $\mathbf{L}(t)$, acted on the fluid by a specific centreline velocity is then found through the integrals
\begin{eqnarray}
\mathbf{F}(t) &=&\mbox{~}\int_{-1}^{1} \mathbf{f}(s,t)\,ds, \label{gelforcerft} \\
\mathbf{L}(t) &=&\mbox{~}\int_{-1}^{1} \mathbf{r}(s,t)\times\mathbf{f}(s,t)\,ds. \label{geltorquerft}
\end{eqnarray}
For a force-free swimmer, the forces and torques from deformation must balance the drag from translation and rotation, thereby determining the swimming velocities. In the following sections we will apply this method to the dynamics of a deforming helical swimmer.

 
\section{Small helix radius deformations} \label{sec:smallamp}

Though the RFT modelling approach can be applied to all   body kinematics relevant to the motion of the deforming helix, the full equations are in general not tractable analytically and reveals little about the physics governing its motion. Hence it is more useful to first consider a limiting configuration. 
Specifically we consider the limit in which the helix radius is small,  $r_{h} \equiv \epsilon r_{h}'$, where $\epsilon\ll 1$ is a small parameter and $r_{h}'$ is fixed, so that the helix is approximately a straight rod with small-amplitude deviations. In this limit, the inextensibility condition can be used to eliminate $\alpha$ through the equation
\begin{equation}
\alpha(t) = 1 -\frac{\epsilon^{2} r_{h}'^{2}(t) k^{2}(t)}{2} + O(\epsilon^{4}), \label{alphasmall}
\end{equation}
thereby making the $r_{h}'$-$k$ configuration space the most practical to work in. In this space, we first identify the different forces and torques on the body before balancing them with rigid-body drag in order to obtain the swimming velocity and the net displacement generated from a given configuration loop.

\subsection{Deformation forces and torques}

Consider the force and torque generated from the deformation of the helix. In the limit of small helix radius, $\epsilon \ll 1$, the deformation velocity becomes
\begin{eqnarray}
\mathbf{V}(t) &=& -s \epsilon^{2} r_{h}'(t) k^{2}(t) \frac{d  r_{h}'}{dt}  \mathbf{\hat{t}} - \epsilon \frac{d r_{h}'}{dt}\mathbf{\hat{n}}  \notag  \\
&& - s \epsilon r_{h}'(t) \left[ \left(1 +\frac{\epsilon^{2} r_{h}'^{2}(t) k^{2}(t)}{2}\right) \frac{d k}{dt}  +\epsilon^{2} r_{h}'(t) k^{3}(t) \frac{d r_{h}'}{dt}\right] \mathbf{\hat{b}}+ O(\epsilon^{4}).
\end{eqnarray}
Hence, using Eqs.~\eqref{gelRFT}, \eqref{gelforcerft}, and \eqref{geltorquerft}, the net force and torque on the fluid from this motion are
\begin{eqnarray}
\mathbf{F} &=& \frac{2\epsilon \zeta_{\perp}}{k^{2}} \left[ \left(k \frac{d r_{h}'}{d t} - r_{h}' \frac{d k }{d t}\right)\sin(k) + k r_{h}' \frac{d k }{d t} \cos(k) \right]\mathbf{\hat{y}}  \notag \\
&&+ 2  \epsilon^{3} k r_{h}'^{2} \frac{d r_{h}'}{dr } (\zeta_{\perp}-\zeta_{\parallel}) \left(k \cos(k) - \sin(k) \right) \mathbf{\hat{y}}  + O(\epsilon^{4}), \label{geldeff}\\
\mathbf{L} &=&-\frac{2 \epsilon  \zeta_{\perp} }{k^{3}} \left[ \frac{d r_{h}' }{d t}  \left(k \sin(k) - k^{2}\cos(k) \right) + r_{h}' \frac{d k }{d t} \left(2 k \cos(k) +(k^{2}-2)\sin(k) \right)\right] \mathbf{\hat{y}}  \notag \\
&& +\frac{r_{h}'^{2} \epsilon^{3}}{k} \left( k \frac{d r_{h}'}{d t} \left[(6 \zeta_{\parallel}-5 \zeta_{\perp}) k \cos(k) +((5-2 k^{2}) \zeta_{\perp}+2(k^{2}-3) \zeta_{\parallel})\sin(k)  \right] \right. \notag \\
 && \left. + \zeta_{\perp} r_{h}' \frac{d k}{d t} \left[4 k \cos(k) + (k^{2}-4) \sin(k) \right] \right)\mathbf{\hat{y}}  +O(\epsilon^{4}). \label{geldefl}
\end{eqnarray}
As anticipated from the symmetry arguments above, both the forces and torques are directed in the $\mathbf{\hat{y}}$ direction. It is also apparent that they are strictly odd functions of the helix radius, $\epsilon r_{h}'$. This is because changing the sign of $\epsilon r_{h}'$ is mathematically equivalent to rotating the helix around its axis, $\mathbf{\hat{x}}$, by $\pi$. Since the rotational symmetry axis, $\mathbf{\hat{y}}$, is perpendicular to $\mathbf{\hat{x}}$, this rotation must also switch the sign of the forces generated by the deformation, thereby requiring the forces and torque to be odd in the helix radius.

\subsection{Rigid-body forces and torques}

In Stokes flow, the net hydrodynamic force and torque arising from rigid-body motion are linearly related to the linear and angular velocity of the body through a symmetric   resistance matrix \cite{Kim2005}.  In the case of the deforming helix, the rotational symmetry around $\mathbf{\hat{y}}$ means only a subset of the coefficients of this matrix are required. Specifically, only the coefficients relating the force and torque in $\mathbf{\hat{y}}$ to the linear and angular velocity in $\mathbf{\hat{y}}$ are needed. The linear relationship in this case can then be written as
\begin{equation}
\left(\begin{array}{c }
F^{y} \\
L^{y}
\end{array} \right) =\left(\begin{array}{c c}
R_{a} &  R_{b}\\
 R_{b} &  R_{c}
\end{array} \right) \left(\begin{array}{c }
U_{r}^{y} \\
\omega_{r}^{y}
\end{array} \right),
\end{equation}
where $\mathbf{U}_{r}$ is the rigid-body translational velocity of the helix, $\boldsymbol{\omega}_{r}$ is the rigid-body angular velocity, and the superscript $y$ denotes the $\mathbf{\hat{y}}$ component of each vector. 
The values of the desired resistance coefficients coefficients, $R_{i}$, can again be found using Eqs.~\eqref{gelRFT}, \eqref{gelforcerft} and \eqref{geltorquerft} by calculating the net hydrodynamic force and torque arising from unit linear and angular velocities. Performing such calculations in the limit of small helix radius we find
\begin{eqnarray}
 R_{a} &=&2 \zeta_{\perp} -\epsilon^{2} (\zeta_{\perp} - \zeta_{\parallel})  r_{h}'^{2} k \left(k- \cos(k) \sin(k) \right) +O(\epsilon^{4}), \\
 R_{b}&=& \frac{\epsilon^{2} r_{h}'^{2} }{4 }  (\zeta_{\perp} - \zeta_{\parallel})\left(4 k +2 k \cos(2 k)- 3 \sin(2 k ) \right)+O(\epsilon^{4}), \\
 R_{c} &=& \frac{2 \zeta_{\perp}}{3 } + \frac{\epsilon^{2} r_{h}'^{2}}{12 k} \left[ 12 k \zeta_{\parallel}  - 18 (\zeta_{\perp} - \zeta_{\parallel}) k \cos(2 k) - 4 (3\zeta_{\perp} - \zeta_{\parallel}) k^{3} \right. \notag \\ 
 &&
 \left. + 3 ((3-2 k^{2}) \zeta_{\perp} - (5-2k^{2}) \zeta_{\parallel}) \sin(2 k)\right] +O(\epsilon^{4}).
\end{eqnarray}
Note that the leading-order values of these coefficients are simply the resistance coefficients of a straight rod, while the $\epsilon^{2}$ terms reflect the small helical deformation. Furthermore, unlike the deformation forces and torques, the resistance coefficients are even functions of helix radius, $\epsilon r_{h}'$, because the drag in $\mathbf{\hat{y}}$ is invariant to rotations of $\pi$ around the helix axis.

\subsection{Instantaneous swimming velocities}

The instantaneous swimming and angular velocities of a deforming helix are then found through balancing the deformation and rigid body forces and torques. Specifically, adding the forces and torques together and setting the result to be zero creates a linear system of equations for both $U_{r}^{y}$ and $\omega_{r}^{y}$, whose solution determines the  swimming velocities of the helix. Using the results above, the instantaneous velocities of the swimmer are
\begin{eqnarray}
U_{r}^{y} &=&-\epsilon \frac{d}{dt} \left(\frac{r_{h}' \sin(k)}{k}\right) - \frac{ \epsilon^{3} r_{h}'^{2} (\zeta_{\perp}-\zeta_{\parallel})}{16 \zeta_{\perp} k^{3}} \left(2 k \frac{d r_{h}'}{d t} f(k) - r_{h}' \frac{d k}{d t} g(k) \right) +O(\epsilon^{4}), \label{rftvelall}\\
\omega_{r}^{y} &=&3 \epsilon \frac{d}{dt} \left(\frac{ r_{h}' \sin(k)}{k^{2}} - \frac{ r_{h}' \cos(k)}{k}\right) +  \frac{3 \epsilon^{3} r_{h}'^{2}}{16 \zeta_{\perp} k^{4} } \left( k \frac{d r_{h}'}{d t} h(k) + 2 r_{h}' \frac{d k}{d t} m(k) \right)  +O(\epsilon^{4}), \label{refovelall}
\end{eqnarray}
where we have defined four functions
\begin{eqnarray}
  f(k) &=&(4 k^{2}-9)(1+2k^{2}- \cos(2 k)) \cos(k)+ k(21-4 k^{2}+15 \cos(2 k))\sin(k), \\
  g(k) &=&(18+53 k^{2} +8 k^{4}) \cos(k) +(19 k^{2}-18)\cos(3 k)  \notag \\
  &&+ 2 k [(4 k^{2}-27)\sin(k)+(2 k^{2} -15) \sin(3 k)], 
\\
 h(k) &=&[(9+9k^{2}+16 k^{4})\zeta_{\perp}-(15+30 k^{2} + 16 k^{4})\zeta_{\parallel} ]\cos(k) \notag \\
 &&+  [(27 k^{2} -9 -2 k^{4})\zeta_{\perp} +(15-30k^{2} +2 k^{4})\zeta_{\parallel}] \cos(3 k) \notag \\
 && -k[(18+3k^{2}-8k^{4})\zeta_{\perp}+(9k^{2} -54+4k^{4})\zeta_{\parallel}]\sin(k) \notag \\
 && + k [(25 k^{2} -54)\zeta_{\perp} + (66-25k^{2})\zeta_{\parallel}] \cos(2 k) \sin(k), \\
 m(k) &=&[(16 k^{4} - 15 k^{2}-9)\zeta_{\perp}+(15+39 k^{2}-40 k^{4})\zeta_{\parallel}]\cos(k) \notag \\
 && +3[(3-7k^{2})\zeta_{\perp}+(7 k^{2}-5)\zeta_{\parallel}]\cos(3 k) \notag \\
 && +2 k[(9-10k^{2}+8k^{4})\zeta_{\perp}+(22 k^{2} -27 - 8k^{4}) \zeta_{\parallel}] \sin(k) \notag \\
 && + 2 k [(27-4k^{2})\zeta_{\perp}+(4 k^{2}-33)\zeta_{\parallel}]\cos(2 k) \sin(k).
\end{eqnarray}
As one may expect, both the translational and angular velocities are  odd functions of the helix radius, $\epsilon r_{h}'$,  again reflecting the relationship between changing the sign of $\epsilon r_{h}'$ and rotating the helix around its axis, $\mathbf{\hat{x}}$.  

Furthermore, the leading-order contributions (order $\epsilon$) in Eqs.~\eqref{rftvelall} and \eqref{refovelall} are exact time derivatives of configuration space parameters and thus cannot generate net motion over a period of deformation. This general result arises from the helix's ability to only swim in one dimension. For one dimensional motion, the net displacement from a given loop in configuration space is given by
\begin{equation}
\left(\begin{array}{c }
\Delta Y \\
\Delta \theta
\end{array} \right) = \oint_{\partial V} \left(\begin{array}{c }
U_{r}^{y} \\
\omega_{r}^{y}
\end{array} \right) \,d t , \label{displacements}
\end{equation}
where $\Delta Y$ is the translational displacement in $\mathbf{\hat{y}}$, $\Delta \theta$ is the rotational displacement around $\mathbf{\hat{y}}$, $t$ is time, and $\partial V$ is a closed loop in configuration space. As the net displacement is a direct integral of the velocity over a closed loop, it is clear that components of the velocity that can be written as exact time derivatives cannot generate motion. Physically this result is due to the leading-order shape of the system. At order $\epsilon$, the body is effectively a straight rod with a prescribed velocity across its length. As a straight inextensible rod has no configurational degrees of freedom, the system cannot break the time symmetry of Stokes flow and so any motion must be the derivative of a periodic function.

As a result, net displacements of the swimming helix can only occur at orders in which the core shape has become helical and can break time-reversal symmetry (order $\epsilon^{3}$ or above). 
It is worth noting that if the drag on the helix was isotropic (ie. $\zeta_{\perp}=\zeta_{\parallel}$), the order $\epsilon^{3}$ terms of the linear velocity become identically zero. Hence, consistent with the current understanding of swimming as low Reynolds numbers, a deforming helix cannot generate net translation in a isotropic drag medium \cite{BECKER2003}. Notably, under the same conditions, the order $\epsilon^{3}$ terms of the angular velocity are non-zero and so net rotation is still possible under isotropic drag. This rotation is a consequence of the anisotropy of the swimmer's shape alone \cite{Koens2016a}.

\subsection{Sensitivity of the net displacements}
 
In order to gain further understanding, we consider how the net displacements, in Eq.~\eqref{displacements}, depend on a particular configuration loop. For this purpose we use the typical dimensions of the deforming gel swimmer from Ref.~\cite{Mourran2017}. The mean (scaled) helix radius of the reported gel swimmer's deformation was $\bar{r}_{h} \approx 0.077$ and it varied by $\Delta r_{h} =0.026$ throughout one cycle. Similarly the mean (scaled) axial length was $\bar{\alpha} \approx 0.903$, and it varied by $\Delta \alpha =0.047$. These dimensions are consistent with the small helix radius assumption of Eqs.~\eqref{rftvelall} and \eqref{refovelall}. An approximate deformation loop of the gel helix can be  parametrised as
\begin{eqnarray}\label{eq:def}
\alpha(t) &=& \bar{\alpha} + \Delta \alpha \sin(t), \label{alpha1}\\
\epsilon r_{h}'(t) &=& \bar{r}_{h} + \Delta r_{h} \cos(t+\phi), \label{radius1}
\end{eqnarray}
where $\phi$ is the phase between the two oscillations and $k$ is  given by Eq.~\eqref{alphasmall}. For $\phi=\pi/2, 3\pi/2$, Eq.~\eqref{eq:def} describes reciprocal motion and therefore no net motion is induced. We note that a different deformation loop would generate different net displacements, as would be expected for any change in a swimming stroke.

\begin{figure}[t]
\includegraphics[width=0.7\textwidth]{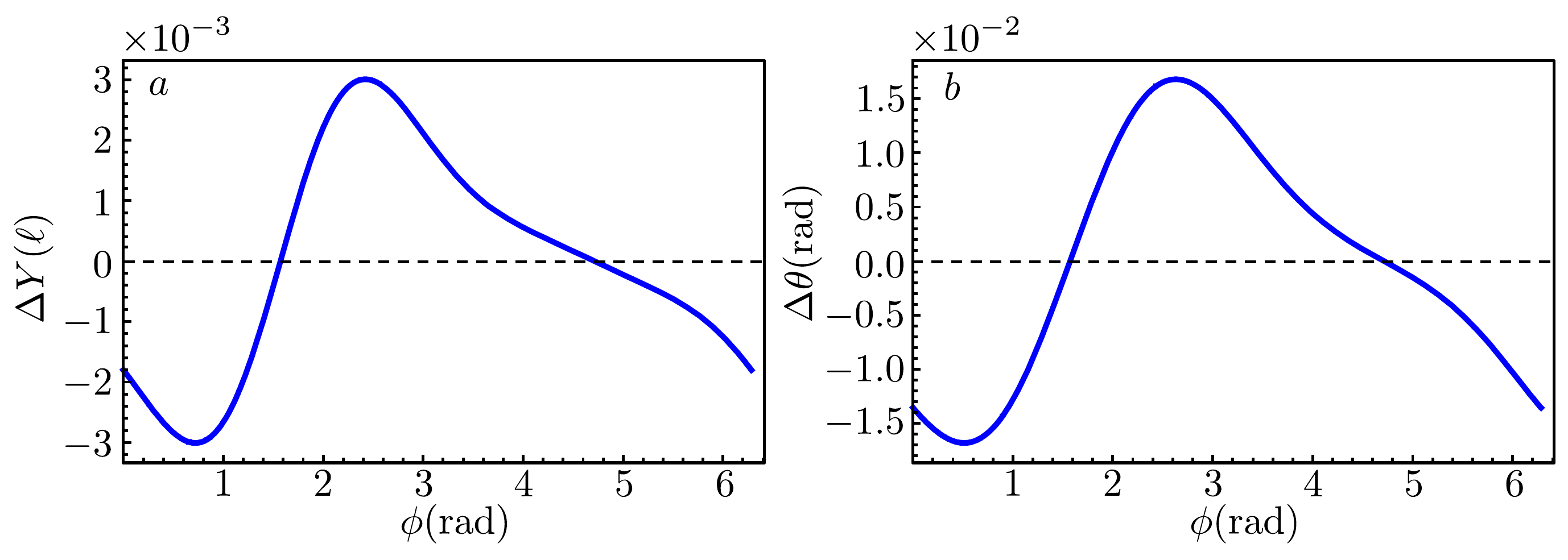}
\caption{Net translational (a) and  rotational (b) displacement generated from a small-radius helix deforming according to Eqs.~\eqref{alpha1} and \eqref{radius1}, as a function of the phase, $\phi$. For this deformation loop, the parameter values are $\bar{\alpha} = 0.903$, $\Delta \alpha =0.047$, $\bar{r}_{h} = 0.077$, and $\Delta r_{h} =0.026$ to reflect the typical dimensions of the experimental gel swimmer in Ref.~\cite{Mourran2017}.
}
\label{fig:phi}
\end{figure}

The configuration loops for $\epsilon r_{h}'$ and $k$ can be inserted into Eqs.~\eqref{rftvelall}, \eqref{refovelall} and \eqref{displacements} to determine the net displacements from one period of deformation. These results are illustrated in Fig.~\ref{fig:phi}. For the loops considered, each displacement shows a different non-sinusoidal dependence on the phase lag, $\phi$. Hence, for typical experimental values, the dynamics of the gel swimmer could vary wildly. On Fig.~\ref{fig:phi} we observe the maximum linear displacement,  $\Delta Y$, to be $\approx 0.003$ while the maximum angle, $\Delta \theta$, to be $\approx 0.015$~rad. Though determined using a RFT model, this maximum rotational displacement is close to the experimentally measured displacement of 0.019~rad~\cite{Mourran2017}. Furthermore, the experiments observed no net translation, in agreement with our prediction that $\Delta Y$ is small. This agreement suggests that some of the experimentally seen motion is governed by similar physics to the deforming helix and so may be understood and optimised using the same principles.

\section{Arbitary deformations} \label{sec:full}

The motion of the helix generated by an arbitrary, periodic configuration loop can also be computed using the RFT formalism. In doing so, it is best to consider the dynamics in the $\alpha$-$r_{h}$ configuration space. As $\alpha$ is the scaled axial length and $r_{h}$ is the scaled helix radius, both variables have clear connections to the physical shape of the helix. Furthermore as $\alpha \in [0,1]$, $r_{h} \in [0,\infty)$ and $k \in [0,\infty)$ and the three parameters are related through the inextensibility condition, Eq.~\eqref{inext}, the physical region of the $\alpha$-$r_{h}$ space is rectangular and easier to visualise than the $r_{h}$-$k$ space, which is bounded by 0 and $r_{h} = 1/k$.

In the $\alpha$-$r_{h}$  configuration space, the instantaneous velocity resulting from a general deformation of the helix can always be formally written as
\begin{equation}
\left(\begin{array}{c }
U_{r}^{y} \\
\omega_{r}^{y}
\end{array} \right) =\mathbf{M}_{r,\alpha} \,\frac{d}{d t}\left(\begin{array}{c }
 r_{h} \\
\alpha
\end{array} \right), \label{linear}
\end{equation}
where the matrix
\begin{equation}
\mathbf{M}_{r,\alpha} \equiv\left(\begin{array}{c c}
M_{r,\alpha}^{Y,r} & M_{r,\alpha}^{Y,\alpha}\\
 M_{r,\alpha}^{\theta,r} &   M_{r,\alpha}^{\theta,\alpha}
\end{array} \right)
\end{equation}
 relates the instantaneous velocities to the rate of change of the helix configuration parameters. The coefficients of this matrix are determined by balancing the forces and torques on the helix. For completeness these coefficients are listed in the appendix in the limit of asymptotically slender filaments, \textit{i.e.}~$\zeta_{\perp} /\zeta_{\parallel}= 2 $. The linear relationship in Eq.~\eqref{linear}, between the deformation variables and the rigid swimming velocity, is the basis for our investigation into general swimming behaviours.

\subsection{Net displacement and motility maps}

For any periodic deformation of the helix, the net linear, $\Delta Y$,  and angular displacement, $\Delta \theta$,  can be written as
\begin{eqnarray}
\Delta Y &=& \oint_{\partial V} \left(M_{r,\alpha}^{Y,r}\frac{d r_{h}}{dl} +  M_{r,\alpha}^{Y,\alpha}\frac{d \alpha}{dl}\right) \,dl, \label{deltaY} \\
\Delta \theta &=& \oint_{\partial V} \left(M_{r,\alpha}^{\theta,r}\frac{d r_{h}}{dl} +  M_{r,\alpha}^{\theta,\alpha}\frac{d \alpha}{dl}\right) \,dl, \label{deltaT}
\end{eqnarray}
where we have expanded Eq.~\eqref{linear}, and parametrised Eq.~\eqref{displacements} by the arc length of the $\alpha$-$r_{h}$ curve in configuration space, $l$. The insensitivity of Eq.~\eqref{displacements} to arbitrary re-parametrisations of time reflects the time invariance of Stokes flow. 
 
The general behaviours and optimal trajectories of the deforming helix swimmer can then be determined by considering how Eqs.~\eqref{deltaY} and \eqref{deltaT} vary for different loops in the $\alpha$-$r_{h}$ configuration space. Typically this is difficult to achieve due to the infinite number of loops possible. However, in this case, the trends and the optimal loops can be determined through a mathematical trick involving the use of Stokes' theorem. Specifically, Eqs.~\eqref{deltaY} and \eqref{deltaT} can be thought of as closed line integrals over vector fields with strengths $\mathbf{M}^{Y} = ( M_{r,\alpha}^{Y,r},M_{r,\alpha}^{Y,\alpha})$ and $\mathbf{M}^{\theta} = ( M_{r,\alpha}^{\theta,r},M_{r,\alpha}^{\theta,\alpha})$, respectively. Stokes' theorem classically states that the value of such closed line integrals is equal to the flux of the curl of this vector field through a surface bounded by the loop. In two dimensions this relationship reduces to Greens' theorem and so Eqs.~\eqref{deltaY} and \eqref{deltaT} can be recast as
 \begin{eqnarray}
\Delta Y &= \displaystyle \iint_{V} \left(\frac{d M_{r,\alpha}^{Y,\alpha}}{d r_{h}} - \frac{d M_{r,\alpha}^{Y,r}}{d \alpha} \right) \,d r_{h} \,d \alpha &= \iint_{V} (\nabla\times\mathbf{M}^{Y})\cdot\mathbf{\hat{x}}_{3}\,d r_{h} \,d \alpha, \label{deltaYA} \\
\Delta \theta &=  \displaystyle \iint_{V} \left(\frac{d M_{r,\alpha}^{\theta,\alpha}}{d r_{h}} - \frac{d M_{r,\alpha}^{\theta,r}}{d \alpha} \right) \,d r_{h} \,d \alpha &= \iint_{V} (\nabla\times\mathbf{M}^{\theta})\cdot\mathbf{\hat{x}}_{3} \,d r_{h} \,d \alpha, \label{deltaTA}
\end{eqnarray}
where $V$ is the area within the loop in configuration space and $\mathbf{\hat{x}}_{3}$ is the unit vector perpendicular to the configuration space. In last equality we have arbitrarily introduced a new orthogonal coordinate $x_{3}$ in which both vector fields have 0 component, ie. $\mathbf{M}^{\theta} =  ( M_{r,\alpha}^{\theta,r},M_{r,\alpha}^{\theta,\alpha},0)$, such that the integrand can be written in terms of the curl.
The identities in Eqs.~\eqref{deltaYA}-\eqref{deltaTA} show that the net translation and rotation of a swimmer only depends on one functional each: $(\nabla\times\mathbf{M}^{Y})\cdot\mathbf{\hat{x}}_{3}$ for translation or $(\nabla\times\mathbf{M}^{\theta})\cdot\mathbf{\hat{x}}_{3}$ for rotation. Plots of these functions over the configuration space are known as {motility maps} or height functions \cite{DeSimone2012, Hatton2015}. They reveal the general behaviour of the system while also providing a quick way estimate the motion from any trajectory. Furthermore, the loops which maximise the values of $\Delta Y$ and $\Delta \theta$, under some length constraint, are just the contours of these curl functions with the same arc length. Hence these maps have been useful in applications of control theory and design optimisation  \cite{Cicconofri2016, Hatton2017, Ramasamy2017}.

\subsection{General displacements of a deforming helix}
 
\begin{figure}[t]
\includegraphics[width=0.85\textwidth]{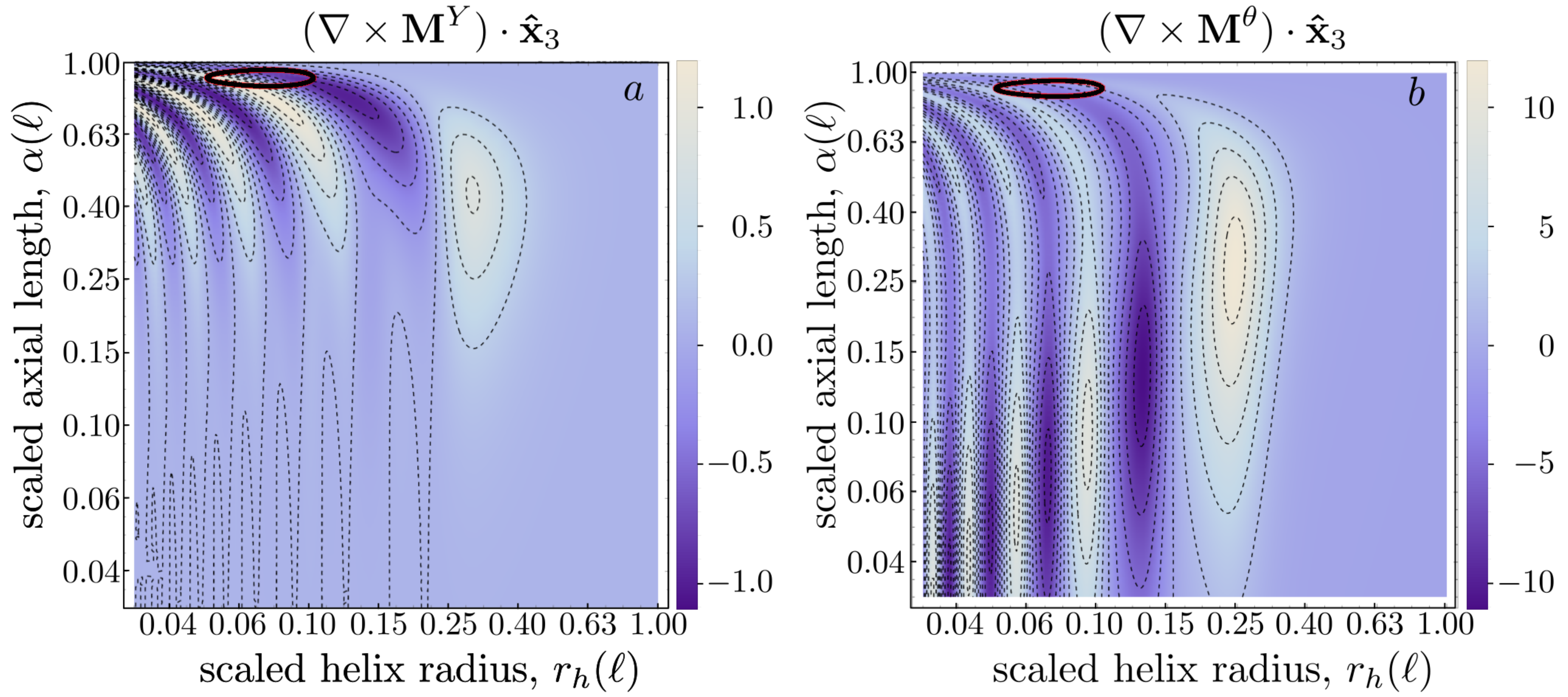}
\caption{Motility maps: Density plots of the functionals $(\nabla\times\mathbf{M}^{Y})\cdot\mathbf{\hat{x}}_{3}$ (a, translation) and $(\nabla\times\mathbf{M}^{\theta})\cdot\mathbf{\hat{x}}_{3}$ (b, rotation) as a function of the scaled helix radius, $r_{h}$, and scaled axial length, $\alpha$.  The dashed lines represent iso-contours of the functionals. The solid black loop represents an approximate path in configuration space taken by the deforming gel swimmer from Ref.~\cite{Mourran2017} (Eqs.~\eqref{alpha1} and \eqref{radius1} when $\phi=0$). 
 Though on a log-log scale, the functions have been scaled such that the total displacement over a loop is the integral of the function within the loop. All lengths have been scaled with respect to $\ell$}
\label{fig:curl}
\end{figure}  
 
In Fig.~\ref{fig:curl}, we plot the motility maps, $(\nabla\times\mathbf{M}^{Y})\cdot\mathbf{\hat{x}}_{3}$ and $(\nabla\times\mathbf{M}^{\theta})\cdot\mathbf{\hat{x}}_{3}$, for a deforming helix swimmer. Though shown on a log-log scale, the functions have been scaled such that the values of $\Delta Y $ and $\Delta \theta$ are still equal to area integrals of the function within the loop. In both panels,  $(\nabla\times\mathbf{M}^{Y})\cdot\mathbf{\hat{x}}_{3}$ and $(\nabla\times\mathbf{M}^{\theta})\cdot\mathbf{\hat{x}}_{3}$ are seen to oscillate, as the helix radius decreases, between positive and negative values of almost equal magnitude but increasing frequency. These oscillations are caused by the changing number of coils within the helix. For an inextensible helix the wavenumber, $k$, is always proportional to $1/r_{h}$. Hence as $r_{h}$ decreases, the number of wavelengths along the length increases and can change the direction of motion. This increasing rate of oscillation suggests that it is better for the swimmer to have a larger helix radius as larger loops that do not cross a zero contour are possible, thereby reducing the sensitivity of the results to small changes while maximising the area within the loop. These oscillations also allow a swimmer of a single chirality to move backwards or forwards when following loops of the same handedness (clockwise or counter-clockwise) and could be harnessed to create loops in which one of the displacements is exactly 0.

Unlike with $r_{h}$, the optimal behaviour of $\alpha$ is seen to be different for translation and rotation. When optimising translation, it is visible that $(\nabla\times\mathbf{M}^{Y})\cdot\mathbf{\hat{x}}_{3}$ is larger when $\alpha$ is closer to 1, while for rotation the maximum of $(\nabla\times\mathbf{M}^{\theta})\cdot\mathbf{\hat{x}}_{3}$ in $\alpha$ is closer to 0.1 and it decreases as $r_{h}$ decreases. This suggests that $\alpha$ can be used to tune the swimmer to be better at either rotating or translating. Note that deforming helices are inherently better at rotating than translation since the maximum of $(\nabla\times\mathbf{M}^{\theta})\cdot\mathbf{\hat{x}}_{3}$ is an order of magnitude larger than $(\nabla\times\mathbf{M}^{Y})\cdot\mathbf{\hat{x}}_{3}$, but the physics causing this difference is yet unclear.
 
The results plotted in  Fig.~\ref{fig:curl} can also be used to estimate the maximum displacements possible from a given loop. For example, let us consider the value of the integral in the rightmost loop shown in each panel, as these are the loops with the largest area. Integrating $(\nabla\times\mathbf{M}^{Y})\cdot\mathbf{\hat{x}}_{3}$ over $r_{h} = [0.2, \infty)$ and $\alpha =[0,1]$, we find the translational displacement of the right most loop to be $\Delta Y \approx 0.08$ in dimensionless units, while similarly integrating $(\nabla\times\mathbf{M}^{\theta})\cdot\mathbf{\hat{x}}_{3}$ over $r_{h} = [0.15, \infty)$ and $\alpha =[0,1]$, we find $\Delta \theta \approx 2.25$~radians. This again reinforces that the deforming helix is better at rotating than translating. Interestingly the net translational displacement of \textit{E.~coli} bacteria per rotation of the flagella ($\sim 100$~Hz), scaled by the  size of the cell body ($\sim 2~\mu$m), is approximately 0.1~\cite{Chattopadhyay2006} and therefore is of a similar magnitude to the maximum $\Delta Y$ obtainable by a deforming helix; it is however one order of magnitude larger than that predicted for the experimental deforming swimmer.
    
Finally the motility maps also explain the sensitivity of the net displacements to the phase, $\phi$, in Fig.~\ref{fig:phi}. In calculating Fig.~\ref{fig:phi} we used  typical dimensions of the experimental gel swimmer from Ref.~\cite{Mourran2017}. The black solid loop plotted in each panel of Fig.~\ref{fig:curl} display such a deformation, as described by Eqs.~\eqref{alpha1} and \eqref{radius1} when $\phi=0$. Clearly these loops encircle regions with both positive and negative amplitude. It is therefore immediately obvious that the net motion observed would depend critically on how this loop interacts with the contours of the motility maps and that loops crossing zero contours would generate less displacement than those aligned.

\section{Breaking additional symmetries} \label{sec:symmetry}

The  deforming gel helix studied experimentally in Ref.~\cite{Mourran2017} was seen to rotate both around and perpendicular to the helix axis. From the theoretical discussion in sect.~\ref{sec:Kinematics} we know that the deforming helix in an unbounded fluid can only translate and rotate in one dimension due to a rotation symmetry. Experimental helices must therefore break this rotation symmetry in order to generate this different behaviour. Hydrodynamically, this symmetry breaking could either result from (i) the presence of nearby surfaces, (ii) external forces such as gravity, or (iii) asymmetric imperfections in the  shape of the helix. 

\begin{figure}[t]
\begin{center}
\includegraphics[width=0.7\textwidth]{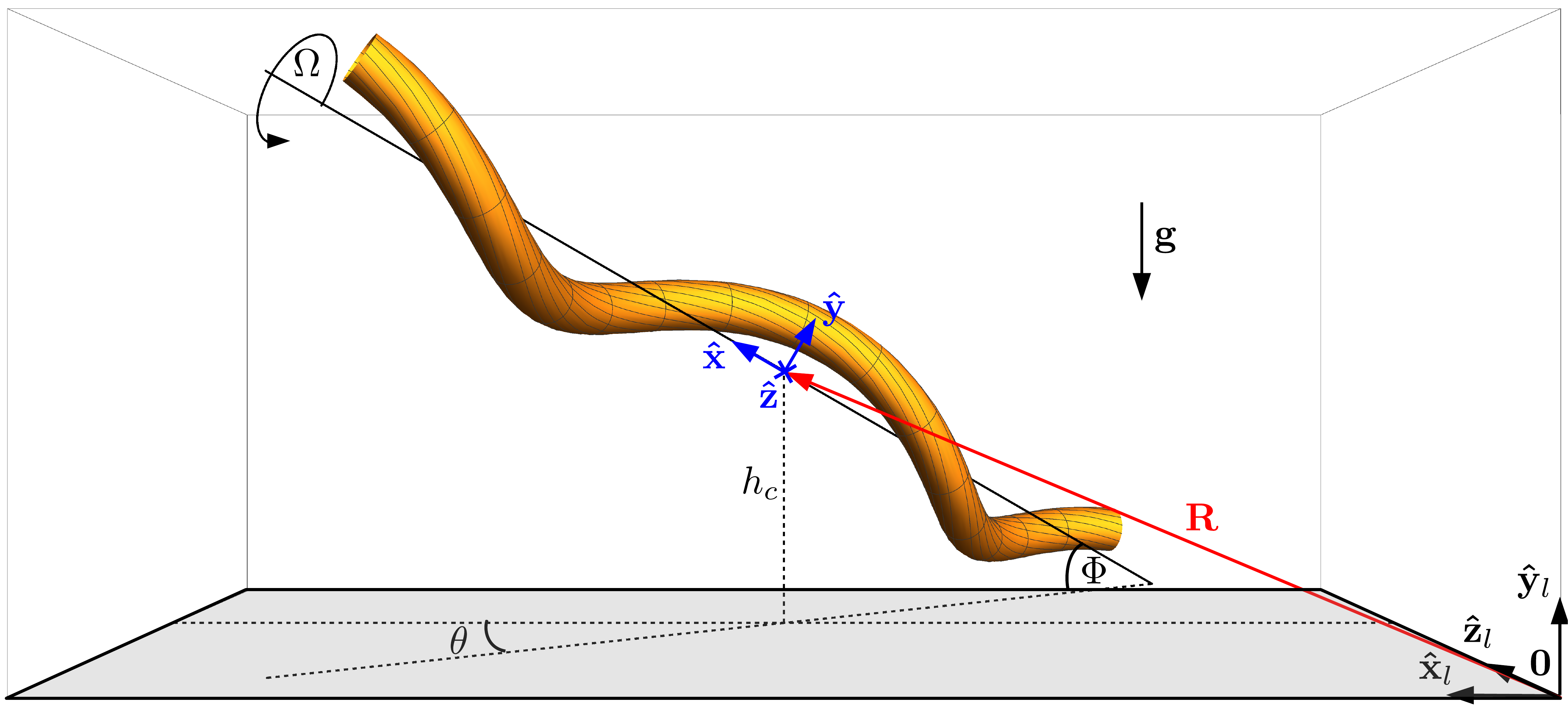}
\caption{Diagram depicting the orientational configuration of the helix in the laboratory frame,  $\mathbf{\hat{x}}_{l}$, $\mathbf{\hat{y}}_{l}$, $\mathbf{\hat{z}}_{l}$. Here $\mathbf{R}$ is the position of the helix in the laboratory frame, $\Omega$ is the rotation angle around the helix axis, $\Phi$ is the angle between the helix axis and the wall, $\theta$ is the angle for rotation around the wall normal, and $\mathbf{\hat{x}}$, $\mathbf{\hat{y}}$, and $\mathbf{\hat{z}}$ are the body frame unit vectors. When relevant, gravity will be defined in the laboratory $-\mathbf{\hat{y}}_{l}$ direction while $h_{c}$ denotes the distance from the centre of the helix to the wall (grey).}\label{fig:angles}
\end{center}
\end{figure}

In this section we consider each of these influences separately and address how they would affect the dynamics of the deforming helix. For simplicity, all three cases are only considered in the asymptotic small helix radius limit, with the configuration loop set by Eqs.~\eqref{alpha1} and \eqref{radius1} (a different configuration loop will, in general, generate different displacements). Furthermore, as breaking the helix symmetry allows full three dimensional motion, it becomes necessary to specify the configuration of the helix in the laboratory frame.
 In what follows the  position of the helix in the laboratory frame is defined by a position vector, $\mathbf{R}$, the rotation around the helix axis, $\Omega$, 
  the angle between the helix axis and the wall, $\Phi$, and the rotation around the wall normal, $\theta$ (see notation in Fig.~\ref{fig:angles}). If 
 $\mathbf{\hat{x}}$, $\mathbf{\hat{y}}$ and $\mathbf{\hat{z}}$ denote unit vectors associated with the body centred coordinate system (consistent with the previous notation), the three dimensional dynamics of the position and orientation of the helix are then determined by \cite{Koens2014}
\begin{eqnarray}
\frac{d \mathbf{R}}{d t} &=& \mathbf{U}_{r} \label{trans}, \\
\frac{d \mathbf{\hat{x}}}{d t} &=&  \boldsymbol{\omega}_{r}\times \mathbf{\hat{x}}, \label{rotx} \\
\frac{d \mathbf{\hat{y}}}{d t} &=&  \boldsymbol{\omega}_{r}\times \mathbf{\hat{y}}, \label{roty} \\
\frac{d \mathbf{\hat{z}}}{d t} &=&  \boldsymbol{\omega}_{r}\times \mathbf{\hat{z}}, \label{rotz}
\end{eqnarray}
and the definition of the three angles   follows from   these unit vectors accordingly.

\subsection{Surfaces}

In experiments, both biological and artificial swimmers are often located close to surfaces either as a result of  their swimming characteristics \cite{Berke2008, Spagnolie2012,Bianchi2017} or because of their differences in density \cite{Zhang2010,Mourran2017}.  Walls therefore play a significant role in the swimmers behaviour, and, in the case of the deforming helix, can break the rotation symmetry if $\mathbf{\hat{y}}$ is not aligned parallel (or anti-parallel) to the wall normal.

The breaking of symmetry in this case occurs because the viscous drag on a body depends on its distance to, and orientation from,  any nearby no-slip surface. The influence of walls on slender filament is, therefore, a difficult theoretical problem and so typically requires the use of numerical techniques \cite{Barta1988, Shum2010, Elgeti2010, Shum2015}. Hence no general version of RFT exists in this case. 
However, if   the filament is oriented perpendicular to the wall normal (\textit{i.e.}~parallel to the wall itself), resistance coefficients have been determined in certain asymptotic limits. In particular, if the pointwise distance between the filament and the wall, $h$, is much less than the body length $\ell$ (\textit{i.e.}~$h\ll\ell$) the resistance coefficients are approximately given by
\begin{align}
\zeta_{t} &\approx \frac{2 \pi \mu}{\log\left(2 h/r_{f}\right)}, \label{wall1}\\
\zeta_{y} &\approx \frac{4 \pi \mu}{\log\left(2 h /r_{f}\right)-1}, \label{wall2} \\
\zeta_{z} &\approx \frac{4 \pi \mu}{\log\left(2 h/r_{f}\right)}, \label{wall3}
\end{align}
where $r_{f}$ is the radius of the filament, $\zeta_{t}$ is the resistance coefficient for drag along the filament axis, $\zeta_{y}$ is the resistance coefficient for drag in the normal direction of the wall and $\zeta_{z}$ is the resistance coefficient for drag in the last direction (see the review in Ref.~\cite{Brennen1977} and   references therein). 

In the small $r_{h}$ limit, the helix is nearly a rod and so, if the angle between the helix axis and the wall, $\Phi$, is small, these resistance coefficients can be used to quantify the locomotion of the helix. Near the wall, the resistive force relationship in Eq.~\eqref{gelRFT}, in this limit, becomes
\begin{equation}
\mathbf{f} \approx \left[\zeta_{t} \mathbf{\hat{t}}\mathbf{\hat{t}} + \zeta_{y} \mathbf{\hat{y}}_{h}\mathbf{\hat{y}}_{h} +\zeta_{z} \mathbf{\hat{z}}_{h}\mathbf{\hat{z}}_{h} \right] \cdot\mathbf{U},
\end{equation}
where $\mathbf{\hat{y}}_{h} = \mathbf{\hat{y}}\cdot\left(\mathbf{\hat{b}}\mathbf{\hat{n}}-\mathbf{\hat{n}}\mathbf{\hat{b}}\right)$, $\mathbf{\hat{z}}_{h} = -\mathbf{\hat{z}}\cdot\left(\mathbf{\hat{b}}\mathbf{\hat{n}}-\mathbf{\hat{n}}\mathbf{\hat{b}} \right)$, $h =\mathbf{\hat{y}}_{l}\cdot\mathbf{R}= \mathbf{\hat{y}}_{l}\cdot\mathbf{r}+h_{c}$ and $h_{c}$ is the height of the centre of the helix  above the wall.  
 This formalism makes it possible to then investigate how the wall affects the motion of a deforming helix. This can be done while maintaining the helix rotation symmetry, if $\mathbf{\hat{y}}$ is aligned with or against the wall normal, or breaking it. When the rotation symmetry is kept, an analytical form of the velocity can be found, however for the full motion (including movement relative to the wall) the solution must be solved numerically.

\subsubsection{Maintaining the helical symmetry}

\begin{figure}[t]
\begin{center}
\includegraphics[width=0.4\textwidth]{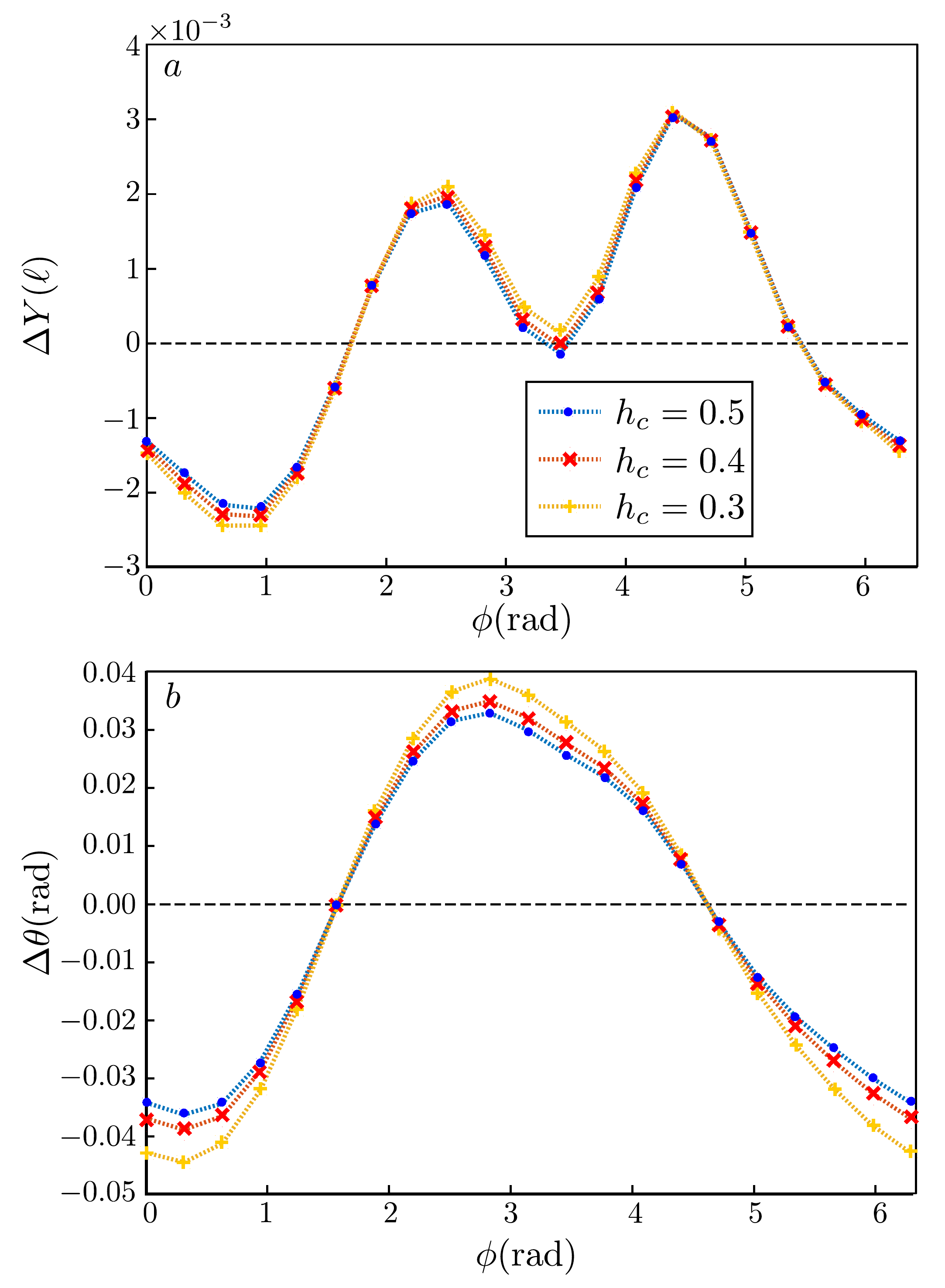}
\caption{Net displacement in translation, $\Delta Y$ (a), and rotation,  $\Delta \theta$ (b),  in the laboratory frame over one period of oscillation using the near-wall resistive force coefficients. Here the cross-sectional radius of the helix is   $r_{f}=0.0375$ corresponding to the effective radius of the gel swimmer in Ref.~\cite{Mourran2017}.}\label{fig:nearwall_disp}
\end{center}
\end{figure}

When the helix symmetry axis, $\mathbf{\hat{y}}$, and the wall normal remain aligned, the helix retains its $\pi$ rotation symmetry. Under these conditions, the motion from any deformation can still only be in $\mathbf{\hat{y}}$ and the leading order linear and angular velocity becomes
\begin{align}
U^{y}_{r} =&\mbox{~}-\epsilon \frac{d}{dt} \left(\frac{r_{h}' \sin(k)}{k} + \epsilon \frac{r_{h}'^{2} (2 k^2 -4 \sin(k)^2 + k \sin(k))}{8 h_{p} k^{2}}\right)  \notag \\
&\mbox{~}- \frac{\epsilon^{3} r_{h}'^{2}}{96 h_{p}^{2} L k^{3}} \left(f_{w} \frac{d r_{h}'}{d t}+  24 r_{h}' g_{w} \frac{d \log(k) }{d t}\right) +O(\epsilon^{4}), \\
\omega^{y}_{r} =&\mbox{~}3 \epsilon \frac{d}{dt} \left(\frac{ r_{h}' \sin(k)}{k^{2}} - \frac{ r_{h}' \cos(k)}{k}\right) + \frac{\epsilon^{2} r_{h}'}{8 h L k} \left(3 n_{w} \frac{d r_{h}'}{d t} + r_{h}' m_{w} \frac{d k}{d t} \right) +O(\epsilon^{3}), 
\end{align}
where we have introduced the four functions
\begin{eqnarray}
f_{w} &=& 12 h_{p}^{2}  k^{3} \cos(k)(2(L-1)\cos(2 k)-(5 L+7) )+48 h_{p}^{2} (L+1) k^{5} \cos(k) \notag \\
&& - 24 h_{p}^{2} (3L+1) k^{4} \sin(k) +96 L \sin^{3}(k)-6 (L+3) k \sin(k) \sin(2k)(3 h_{p}^{2} +2 L ) \notag \\
&&+ k^{2} (2 \sin(k) (4 L (L-3) +3 h_{p}^{2} (11 L+ 9)) + \sin(3 k) (4 L (L+1)  - 9 h_{p}^{2} (L-5))), \\
g_{w} &=& -3 k^{3} \cos(k) (8 L(L+1) -3 h_{p}^{2} (19L +17) +\cos(2 k) (3 h_{p}^{2}(7 L-19) -4 L (L+1))) \notag \\
&& +2 k^{2} \sin(k) (\cos(2 k ) (27 h_{p}^{2} (L-5) - L (11 L +29)) -L(L+1) -27 h_{p}^{2}(5 L +7) )  \notag \\
&&  +36 h_{p}^{2} (3 L +1) k^{5} \cos(k)-144 L \sin^{3}(k) + 18 h_{p}^{2} k^{4} (2 (L+1) \sin(k) -(L-1) \sin(3 k))  \notag \\
&& +36 k \cos(k) \sin(k)^2 (2 h^{2} (L-1)^{2}(L+3)+L(L+6)),\\
n_{w} &=&  12(2+k^{2}) -2 \cos(2 k)(12-18 k^{2} + k^{4}) + k \sin(2 k) (13 k^{2} - 48), \\
m_{w} &=& 6 \cos(2 k) (24 - 48 k^{2} + 7 k^{4}) + 4(k^{6} -9 k^{4} -36) + 3 k \sin(2 k)(96-49 k^{2} +2 k^{4}),
\end{eqnarray}
$h_{p} = h_{c} (L-1)$ and $L=\log(2 h_{c}/r_{f})$.
The above expansion shows that, in the presence of a wall, the net leading-order angular velocity becomes of order $\epsilon^{2} r_{h}'^{2}$ while the leading-order linear velocity remains at $\epsilon^{3} r_{h}'^{3}$. The presence of a surface therefore causes a significant increase in the angular velocity while maintaining a similar magnitude linear velocity. 

Significantly, the $\sim \epsilon^{2} r_{h}'^{2}$ nature of the angular velocity indicates that the direction of rotation is independent of whether the symmetry axis is parallel or anti-parallel to the wall normal. Hence, since the free motion must be an odd function of $\epsilon r_{h}'$, this motion arises uniquely from the hydrodynamic interactions of the helical body with the wall. 

The net displacement of the deforming swimmer can be determined by substituting these velocities into Eqs.~\eqref{trans}-\eqref{rotz}. In doing so, care is needed to account for the changing height of the swimmer above the wall, $h_{c}$. We show in Fig.~\ref{fig:nearwall_disp} the net linear and angular displacement for different initial heights. As the body gets closer to the wall, the magnitudes of both the net translation and rotation increase. Hence walls could be exploited to enhance the  net motion of a deforming helix. 

 However, from our predictions, we see that this rate of increase remains small as $h_{c}$ decreases. This is a consequence of the logarithmic dependence of the resistance coefficients, Eqs.~\eqref{wall1}, \eqref{wall2} and \eqref{wall3},  on the separation from the wall.  This logarithmic dependence will  change as the distance from the wall becomes less than the thickness of the body, $r_{f}$, and the systems enters the lubrication limit \cite{Kim2005}. In that case, the local drag on the cylinder is known to increase as $\sim 1/\sqrt{h-r_{f}}$ \cite{Cardinaels2015} and therefore the presence of the wall would lead to  a larger effect.

\subsubsection{Breaking the helical symmetry}

When $\mathbf{\hat{y}}$ is not aligned with the wall normal, the presence of the surface breaks the helix's $\pi$ rotation symmetry. Theoretically, such configurations can be achieved by rotating $\mathbf{r}$ around its axis by an angle $\Omega_{i} \neq 0,\pm\pi$ 
 before placing it near a wall. This configuration captures a deforming helix `lying' near the bottom wall at an arbitrary initial orientation.

\begin{figure}[t]
\begin{center}
\includegraphics[width=0.8\textwidth]{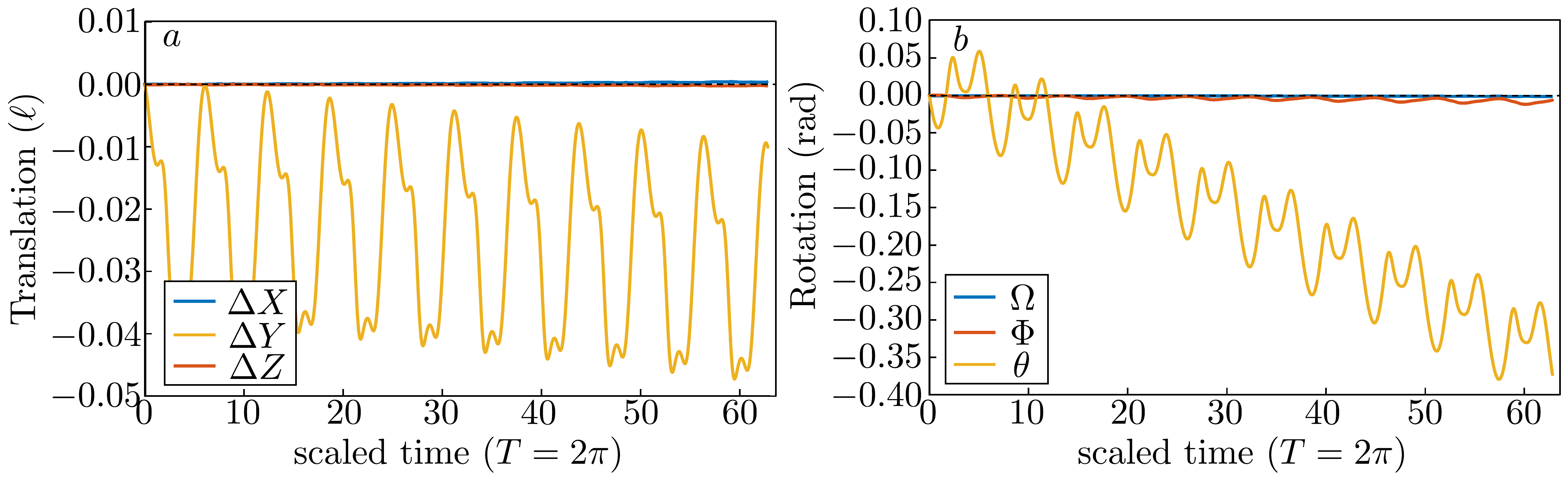}
\caption{Net laboratory-frame  displacement in translation (a) and rotation  (b) over ten periods of oscillation using the near-wall resistive force coefficients. This helix initially has an orientation of $\Omega_{i}=\pi/2$ and a distance from the wall of $h_{c}=0.3$. The filament radius is $r_{f}=0.0375$ and $\phi=0$. The three angles are defined geometrically in the caption of Fig.~\ref{fig:angles}.}
\label{fig:offaxis}
\end{center}
\end{figure} 

In Fig.~\ref{fig:offaxis} we show the (numerically determined) laboratory-frame orientation and position of a helix with an initial configuration $\Omega_{i}= \pi/2$, and $h_{c}=0.3$. Configurations with different initial values for $\Omega_{i}$  exhibit qualitatively similar behaviour. The results show that breaking the helix rotation symmetry leads to motion in all directions. However, of all the components of translations and rotations, motion in the $\mathbf{\hat{y}}_{l}$ direction still dominates. After $\theta$, $\Phi$ is the next largest angle. Since $\Phi$ represents the angle between the helix axis and the wall (Fig.~\ref{fig:angles}), its change indicates that the helix is also slowly aligning its helix axis, $\mathbf{\hat{x}}$, with the wall normal. There is however little rotation around said helix axis or displacement in $\mathbf{\hat{x}}_{l}$ and $\mathbf{\hat{y}}_{l}$. Therefore, though the wall can break the rotational symmetry, in our model it only tends to weakly increase the net displacements over a period.

\subsection{Gravity}

The presence of gravity, acting as an external force, can also generate motion and break the helix's rotation symmetry. Though not actually related to the deformation of the helix, this motion occurs simultaneously and so can play part in the observed dynamics. The significance of the gravitational motion is determined by the ratio of the gravitational and viscous forces. Often called the Archimedes number, $Ar$, this ratio is defined as
\begin{equation}
Ar = \frac{g (\rho_{b}- \rho_{f}) V T}{\mu \ell^{2}},
\end{equation}
where $g = 9.8$~m~s$^{-1}$  is the gravitational acceleration, $\rho_{b} $ is the mass density of the helix, $\rho_{f}$ is the density of water, $V$ is the volume of the swimmer, $\mu$ is the dynamic viscosity of the fluid (water in the experiments) and $T$ is the period of the shape oscillation. For the gel helix swimmer studied in Ref.~\cite{Mourran2017}, we have $Ar \approx 0.058$.

The velocities generated by gravity can be found by balancing the gravitational (buoyancy) force with the drag force and torque from rigid body motion. This is best done in an infinite fluid. In the body frame the gravitational acceleration is given by
\begin{equation}
\mathbf{g} =- g \left\{ \sin(\Phi), \cos(\Phi) \cos(\Omega), -\cos(\Phi) \sin(\Omega)\right\}.
\end{equation}
  Balancing the resulting gravitational force with the viscous drag, and expanding in small  helix radius, $\epsilon r_{h}'$, we obtain the leading-order velocity due to  gravity as
\begin{eqnarray}
U^{x}_{r} &=& Ar \frac{ \sin(\Phi)}{2 \zeta_{\parallel}} +O(\epsilon),\\
U^{y}_{r} &=& -Ar \frac{\cos(\Phi) \cos(\Omega)}{2 \zeta_{\perp}} +O(\epsilon), \\
U^{z}_{r} &=& Ar\frac{ \cos(\Phi) \sin(\Omega)}{4 \zeta_{\perp}\gamma}\left\{\left[\cos(2k)-1-2k^2\right](3-k^2)-6 k \sin(2k)\right\} +O(\epsilon),\\
\omega^{x}_{r} &=& Ar \frac{k  \sin(\Phi) }{4 \zeta_{\perp} \zeta_{\parallel} \gamma }\left\{2 (\zeta_{\perp}-\zeta_{\parallel})\left[2+\cos(2k)\right]k^{2} -2(\zeta_{\perp}-\zeta_{\parallel}) k^{4} \right\}  \notag \\ 
&&+ Ar \frac{k  \sin(\Phi) }{4 \zeta_{\perp} \zeta_{\parallel} \gamma }\left[ 6 \zeta_{\perp} \sin^{2}(k)-3(2 \zeta_{\perp}-\zeta_{\parallel}) k \sin(2k)\right] +O(\epsilon) \\
\omega^{y}_{r} &=&  -  Ar~\epsilon^{2}r_{h}'^{2}\frac{3 \cos(\Phi) \cos(\Omega) (\zeta_{\perp}-\zeta_{\parallel})}{16 \zeta_{\perp}^{2}} [2 k (2+\cos(2k))-3 \sin(2k)]  +O(\epsilon^{3}),\\
\omega^{z}_{r} &=& Ar \frac{2   k \cos(\Phi) \sin(\Omega) }{2 \zeta_{\perp} \gamma} [\sin(k)- k \cos(k)]+O(\epsilon),
\end{eqnarray}
where we used the function $\gamma$
\begin{eqnarray}
\gamma =  k^4 + 3 k \sin(2 k) - (2+\cos(2k)) k^{2} - 3 \sin^{2}(k).
\end{eqnarray}
This result shows that gravity generates leading order motion in every direction except for the rotation around $\mathbf{\hat{y}}$, which occurs at order O($Ar\,\epsilon^{2}$). Therefore, if the value of $Ar$  was  large enough, gravity could create significant motion in the directions excluded by the helical symmetry.

\subsection{Imperfect helix shapes}
 \begin{figure}[t]
\begin{center}
\includegraphics[width=0.8\textwidth]{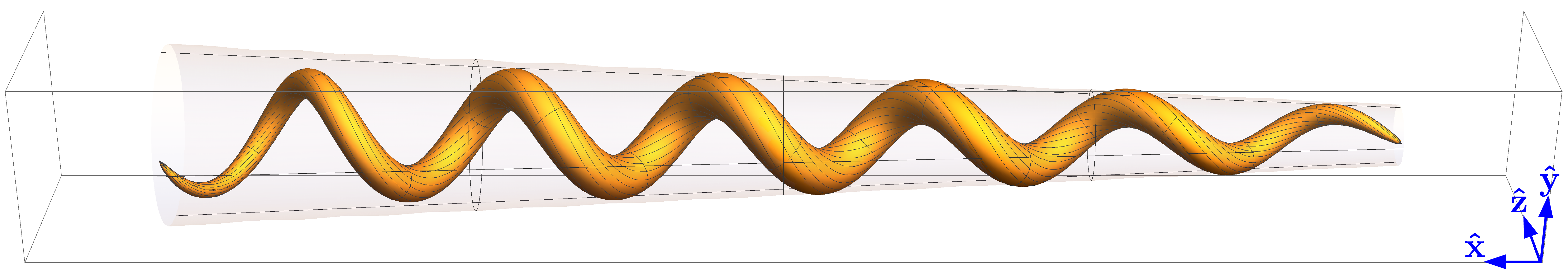}
\caption{Configuration of a conical helix in its body frame.}\label{fig:deformed}
\end{center}
\end{figure}

Defects in the shape of the helix  may also break the $\pi$ rotation symmetry and so generate full three-dimensional motion, the specific form of the defect ultimately determining the resultant motion. As the range of defects possible is effectively infinite, in this section we focus on the behaviour of a small conical defect  causing the helix radius to monotonically increase from one side of the helix to the other (see illustration in Fig.~\ref{fig:deformed}).  Mathematically the centreline of the conical helix can be written as
\begin{equation} \label{defect}
\mathbf{r}(s,t) = \left\{ \alpha(t) s, \epsilon r_{h}'(t)(1+\xi s) \cos(k(t)s), \epsilon r_{h}'(t)(1+\xi s)  \sin(k(t)s) \right\},
\end{equation}
where $\xi$ is a small parameter, so that $\alpha^{2} +(1+\xi s)^2 r_{h}'^{2} k^{2} \approx 1$. This small change is enough to break the symmetry of the shape and so enable fully three dimensional motion.

\begin{figure}[t]
\begin{center}
\includegraphics[width=0.8\textwidth]{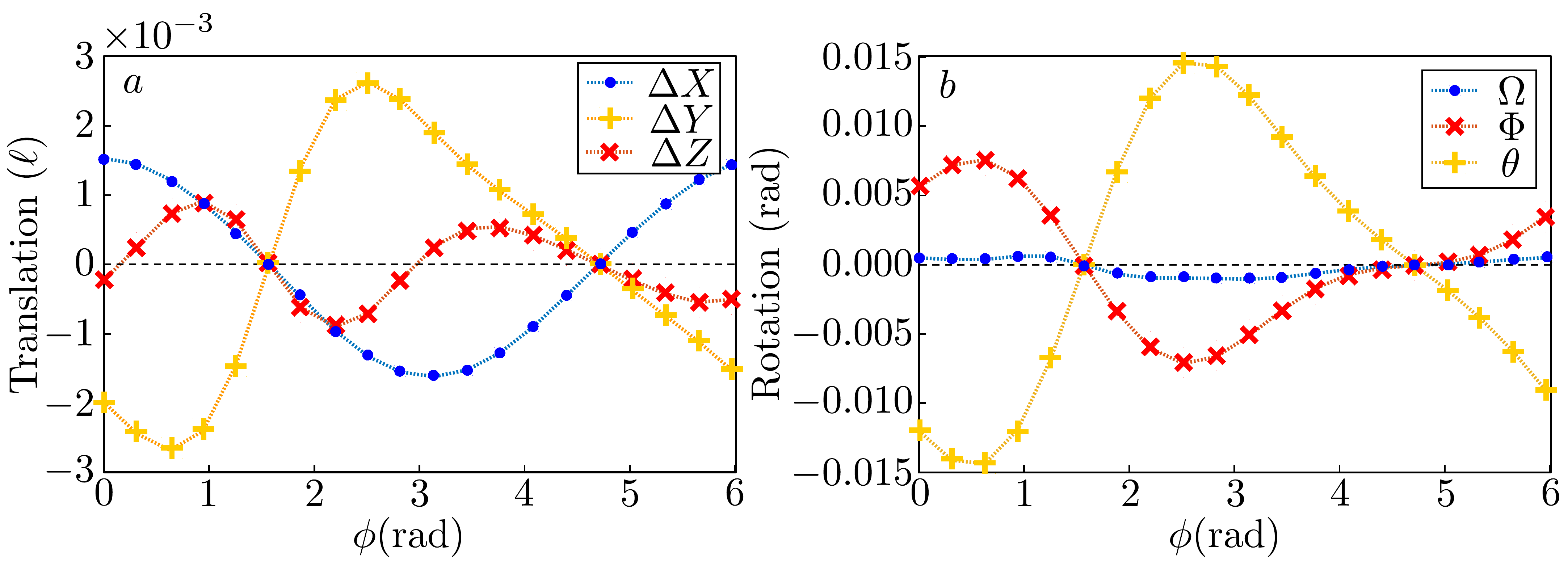}
\caption{Net laboratory-frame  displacement in translation (a) and in rotation  (b) over one period of oscillation for a conical helix with $\xi=0.1$.}\label{fig:deformedmotion}
\end{center}
\end{figure}

We plot the net displacement after one period of deformation as a function of the phase, $\phi$, in Fig.~\ref{fig:deformedmotion}. In these plots $\xi=0.1$ and the configuration loop is again prescribed by Eqs.~\eqref{alpha1} and \eqref{radius1}. For the conical helix, it is clear that  all displacements are non-zero and non-trivially depend on the value of $\phi$. Defects in the shape can therefore generate displacements in all possible directions, enabling full three-dimensional motion.

\section{Discussion}

Helices are iconic shapes in the mechanics of small-scale locomotion, with many types of swimmers rotating rigid helices to generate thrust. However, deforming helices can also create non-trivial motion. Inspired by the range of systems displaying motion from the deformation of a helix, we considered in this paper how an inextensible helix can propel at low Reynolds number  by changing its helix angle, helix radius and wavenumber uniformly across its body. We show that these deforming helices can create non-reciprocal strokes and, in an infinite fluid, can only move in one dimension, namely along a direction perpendicular to the helix axis. In the limit of small helix radius, the velocity of these swimmers was shown to be an odd function of the helix radius and could only generate displacement scaling as the cube of the  helix radius. For helices with  dimensions similar to the deforming gel swimmer of Ref.~\cite{Mourran2017}, the value of the net displacements over one period of oscillation depended strongly on the configuration loop taken but could be of a similar order of magnitude to the displacements found in the experiment.

The net translation and rotation of general unbounded deforming helices were then explored using {motility maps}. These maps use Stokes' theorem to relate the closed line integral representation of the net linear and angular displacement of the swimmer to an area integral of a single scalar function. Inspection of this function revealed the general trends of the swimmer, and clearly showed the optimal loops (the contours). For the deforming helix, the changing number of wavelengths along the helix causes these functions to oscillate. As a result larger displacements were possible at larger helix radii for both the translational and angular displacement. The motility map functions further revealed that the rotational displacement is optimised at small axial lengths while translational displacements is optimised at large axial lengths. 

Using the small-radius approach, we next considered how walls, gravity and a conical shape defect could break the rotational symmetry of the deforming helical swimmer and how it would change the behaviour seen. In the presence of a no-slip surface, the net displacement of the helix, in the direction perpendicular to the wall, weakly increased the closer to the wall it sat.  However, the presence of the surface generated negligible motion in any other direction. On the other hand, gravity was observed to be able to create motion in every direction and so, provided buoyancy is relevant, could be a generator of the motions that appear to break the rotational symmetry in the experiments \cite{Mourran2017}. Similarly, a conical shape defect on the swimmer was also shown to generate significant motion in every direction. 

Through our  exploration, we have created a theoretical basis to understand many of the motions generated by deforming helices. These results in turn can then be used to guide the design of other artificial deforming helix systems, similar to the gel helix from Ref.~\cite{Mourran2017}. These predictions could be further improved through more accurate hydrodynamic models or considering how gel extensibility affects the dynamics. This investigation and optimisation could also lead to the design of tunable artificial swimmers, by introducing how the swimmers react to external fields to our model. Finally, the combination of this model with deformation models for the body of a \textit{Spirochaetes} or the polymorphic transitions of a bacterial flagellum, could also help improve our understanding into how biological systems employ deforming helices to generate motion and their efficiency in doing so.
 
\paragraph*{Acknowledgements}
This project has received funding from the European Research Council (ERC) under the European Union's Horizon 2020 research and innovation programme  (grant agreement 682754 to EL).
 We also gratefully acknowledge support from the DFG within the priority program SPP 1726 on €œMicroswimmers from Single Particle Motion to Collective Behaviour (AM and MM).

\paragraph*{Author Contributions}
LK and EL developed the theoretical model and ran the numerical simulations.
HZ, MM and AM provided the descriptions and understanding of the experimental system.
All authors contributed to the interpretation of the results and to the final manuscript.

\appendix
\section{$\mathbf{M}_{r,\alpha}$ coefficients}

In this appendix we list the coefficients of the $\mathbf{M}_{r,\alpha}$ as found using RFT. Assuming $\zeta_{\perp} = 2\zeta_{\parallel}$ these coefficients become
 \begin{eqnarray}
\frac{\beta M_{r,\alpha}^{Y,r}}{8} &=&-3 \left(\alpha ^2-1\right) \left[\alpha ^4 \left(k^2-4\right)+\alpha ^2 \left(2-3 k^2\right)-12\right] \cos (3 k) \notag \\
&& -4 k \left[5 \alpha ^6 \left(k^2-3\right)+12 \alpha ^4
   \left(k^2-1\right)+3 \alpha ^2 \left(5 k^2-4\right)+39\right] \sin (k) \notag \\
   &&+\left[-6 \left(2 \alpha ^6-3 \alpha ^4+7 \alpha ^2-6\right)+8 \alpha ^2 \left(\alpha ^4+4 \alpha
   ^2+3\right) k^4\right] \cos (k) \notag \\
   &&+\left(-45 \alpha ^6-60 \alpha ^4+57 \alpha ^2+48\right) k^2 \cos (k) +12 \left(\alpha ^6-2 \alpha ^4+2 \alpha ^2-1\right) k \sin (3 k), \\
 \frac{r_{h} \beta M_{r,\alpha}^{Y,\alpha}}{16 \alpha} &=& \left[3 \alpha ^2 \left(4 r_{h}^2+8 k^2+9\right)+3 \left(r_{h}^2+6\right)+\alpha ^4 \left(8 k^2-45\right)\right]  \cos (k)+3 \left(1-4 \alpha ^2\right) r_{h}^2 k \sin (3 k) \notag \\
 &&-k \left(20
   \alpha ^4+12 \alpha ^2 \left(5 r_{h}^2+1\right)+33 r_{h}^2\right) \sin (k)-3 \left(\alpha ^4+\alpha ^2 \left(4 r_{h}^2-3\right)+r_{h}^2+2\right) \cos (3 k), \notag \\ \\
\frac{\beta M_{r,\alpha}^{\theta,r}}{24 k \alpha} &=& \left(\alpha ^2+4\right) \alpha ^2 k \left(8 k^2-21\right) \sin (k)+3 \left[\left(\alpha ^2-1\right)^2+4 \left(\alpha ^4+8 \alpha ^2+7\right) k^2\right] \cos (k) \notag \\
&& +3 k \left(8
   k^2-29\right) \sin (k)-3 \left(\alpha ^2-1\right)^2 \cos (3 k)-\left(\alpha ^4-4 \alpha ^2+3\right) k \sin (3 k), \\
  \frac{r_{h} \beta M_{r,\alpha}^{\theta,\alpha}}{48} &=& \left(3 \alpha ^2-2\right) r_{h}^2 k \cos (3 k)+k \left[12 \alpha ^4+76 \alpha ^2+\left(2-3 \alpha ^2\right) r_{h}^2-24\right] \cos (k)\notag \\
  && +\left\{\alpha ^2 \left[-21 \alpha ^2+8
   \left(\alpha ^2+3\right) k^2-65\right]+22\right\} \sin (k)-\left(\alpha ^4-3 \alpha ^2+2\right) \sin (3 k), \\
  \beta &=& 3 \left(\alpha ^2-8\right) \left(\alpha ^2-1\right)^2+32 \alpha ^2 \left(\alpha ^2+3\right)^2 k^4-24 (\alpha -1) (\alpha +1) \left(\alpha ^4-9 \alpha ^2+24\right) k^2 \notag \\
  &&+8
   \left(\alpha ^2-1\right) k \left[\alpha ^4 \left(4 k^2-3\right)+3 \alpha ^2 \left(4 k^2-11\right)+24\right] \sin (2 k) \notag \\
   &&+48 \alpha ^2 \left(\alpha ^4+10 \alpha ^2-11\right)
   k^2 \cos (2 k)-3 \alpha ^2 \left(\alpha ^4-10 \alpha ^2+17\right) \cos (4 k)+24 \cos (4 k). \notag \\
 \end{eqnarray}
 These coefficients are used to calculate the motility maps in the main text. Note that in order to take the derivative of these coefficients $k$ must be eliminated from the equations using the inextensibility condition.

\bibliographystyle{unsrt}
\bibliography{library}
\end{document}